\documentclass[aps,twocolumn,amsmath,amssymb,superscriptaddress,floatfix]{revtex4}
\usepackage{amsmath,amssymb,amsfonts,mathrsfs}
\usepackage{epsf}
\usepackage{psfrag}
\usepackage{graphicx}
\usepackage{color}
\begin{document}

\author{Wade DeGottardi}
\affiliation{Department of Physics, University of Illinois at
Urbana-Champaign, 1110 W.\ Green St.\ , Urbana, IL  61801-3080, USA}

\author{Tzu-Chieh Wei}
\affiliation{Department of Physics and Astronomy, University of
British Columbia, Vancouver, BC V6T 1Z1, Canada}

\author{Victoria Fernandez}
\affiliation{Departamento de F\'{\i}sica, Universidad Nacional de La
Plata and Instituto de F\'{\i}sica  La Plata, CONICET, Argentina}

\author{Smitha Vishveshwara}
\affiliation{Department of Physics, University of Illinois at
Urbana-Champaign, 1110 W.\ Green St.\ , Urbana, IL  61801-3080, USA}

\date{\today}

\title{Accessing nanotube bands via crossed electric and magnetic fields}

\begin{abstract}
We investigate the properties of conduction electrons in
single-walled armchair carbon nanotubes in the presence of mutually
orthogonal electric and magnetic fields transverse to the tube's
axis. We find that the fields give rise to an asymmetric dispersion
in the right- and left-moving electrons along the tube as well as a
band-dependent interaction. We predict that such a nanotube system
would exhibit spin-band-charge separation and a band-dependant
tunneling density of states. We show that in the quantum dot limit,
the fields serve to completely tune the quantum states of electrons
added to the nanotube. For each of the predicted effects, we provide
examples and estimates that are relevant to experiment.
\end{abstract}

\maketitle

A highlight of graphene-based nanotubes is the presence of a band
(valley) quantum degree of freedom which offers a new facet to a
range of strongly correlated low-dimensional phenomena. In armchair
nanotubes, the bands, which owe their existence to the two-atom basis
of the underlying honeycomb lattice, form a pair of gapless linearly
dispersing modes~\cite{dresselhaus}. While a direct consequence of
their presence is the measurable quantum conductance of $4e^2/h$,
richer effects can arise, for instance, in field-induced orbital
moments~\cite{mceuen1}, Coulomb-blockade profiles~\cite{bockrath} and
in Kondo physics~\cite{kondo}. An ability to precisely control the
band sector would not only shed light on these effects but also
access other band-dependent physics. For instance, in the recently
reported nanotube Mott phase~\cite{MottBockrath}, the nature of this
ordered phase could be probed by tuning inter-band interactions. In
the Luttinger liquid phase typical of interacting one-dimensional
systems, spin and charge sectors have been shown to decouple in
quantum wires~\cite{Yacoby05}; nanotubes could extend such
fractionalization into yet another sector. Recently, attention has
turned towards ``valleytronics'' as a potential application for
quantum information and quantum devices in the parent system of
graphene~\cite{valleytronics}; the ability to manipulate spin and
band quantum states in nanotubes would offer an attractive
alternative. However, save for some exceptions~\cite{mceuen1}, the
band degree of freedom has thus far remained relatively resilient to
any controlled coupling.

\begin{figure}
    \includegraphics[bb = 60 240 360 540,width=57mm]{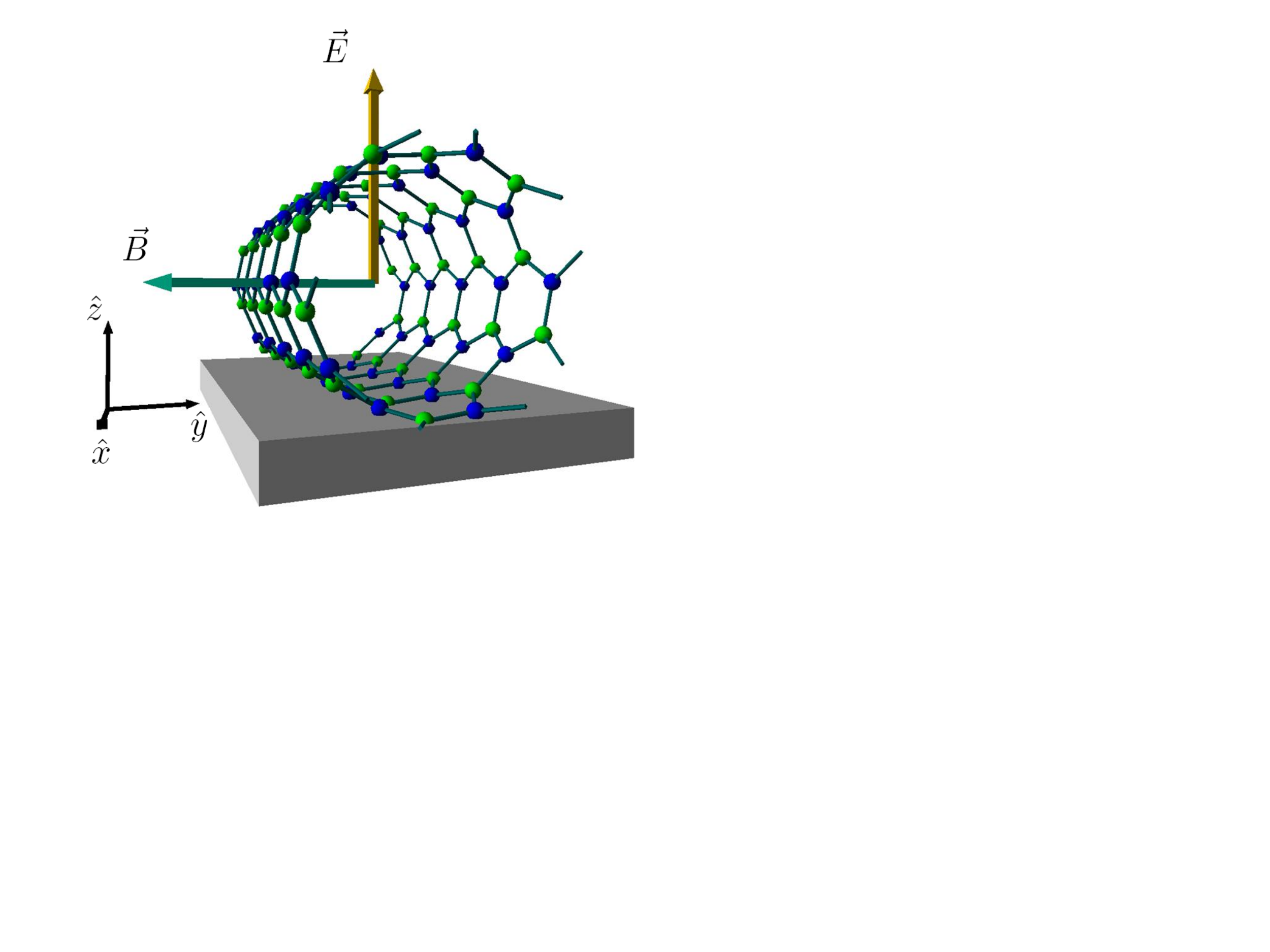}
     \caption{(Color online) An armchair carbon nanotube in the presence of transverse magnetic (pointing in the $-\hat{y}$-direction) and electric fields (pointing in the $\hat{z}$-direction). The carbon atoms belonging to the A
     and B sublattices are indicated by dark and light shading, respectively.}
     \label{fig:setup}
\end{figure}

In this Letter, we explore the possibility of coupling to the band
sector via applied transverse fields. Electric and magnetic fields
individually have been shown to alter the low energy properties of
conduction electrons~\cite{bellucci, novikov}. Here, building on our
prior work~\cite{us}, we find that the transverse field configuration
shown in Fig.~\ref{fig:setup} in which both fields are present offers
a powerful means of accessing the band sector. The fields
 break time-reversal, particle-hole and band symmetries, and in fact, render the
 dispersion asymmetric in the right- and left-moving electrons.
Furthermore, the effective Coulomb interaction is modified from its
field-free form to include couplings between the charge and band
sectors. As a consequence, nanotubes in transverse fields could
provide a realization of a tunable asymmetric Luttinger
liquid~\cite{trushin, naon, gil}. Most pertinently, the system
displays spin-band-charge separation, acts as a band-selector for
electrons tunneling into the nanotube and exhibits a variety of
tunable quantum dot shell filling configurations.

The setup of interest, as shown in Fig.~\ref{fig:setup}, consists of
an armchair nanotube lying along the $x$-axis subject to an applied
magnetic field $\vec{B} = -B\hat{y}$ and electric field $\vec{E} =
E\hat{z}$.
 For a tube of radius $R = 3 n a_c/2 \pi$, where $n$
is the chiral index and $a_c \approx 0.142$ nm is the carbon-carbon
bond length, the effect of the fields can be characterized by the
parameters $b = \sqrt{3} B |e| R^2/\hbar$ and $u = |e| E R / t$,
where $t \approx 2.7$ eV~\cite{dresselhaus} is the electron hopping
strength between neighboring sites. Near half-filling, the field-free
low-energy electronic dispersion,
 which can be traced
to bonding and anti-bonding states of  the underlying graphene sublattices,
 consists of four degenerate gapless linear modes each having a Fermi velocity $v_F = \frac{3
t a_c}{2 \hbar} \approx 8 \times 10^5$ m/s. These modes correspond to
right- and left-movers (indexed by $r=\pm=R/L$) at two Fermi points
(indexed by $\alpha=\pm$) which define the band degrees of freedom.

In this Hall-type setup, classically, each of the fields causes a
spiralling motion of electrons, and thus for fixed kinetic energy,
reduces their linear velocity along the tube's axis. Moreover, the
non-vanishing Poynting vector $\vec{E}\times\vec{B}$ affects right-
and left-movers differently. Indeed, a detailed perturbative band
structure calculation~\cite{us} for large enough tubes ($10\lesssim n
<\pi/u$) shows that the right- and left-moving speeds are given by
$v_{r} \approx v_F \big(1 - \frac{1}{3} b^2 - \frac{n^2}{\pi^2} u^2
\pm \frac{n}{\pi} b u \big)$. The spectrum remains
gapless~\cite{gap}, but the two Fermi points reside at energies
differing by $\Delta_F\approx 2 \pi t b u / (\sqrt{3} n)$, indicating
a mild band-degeneracy breaking even at the band structure level. The
corresponding low-energy non-interacting asymmetric Hamiltonian can
be expressed as
\begin{equation}
H_0 = -i \hbar \sum_{r \alpha \sigma} \int dx \ r v_r \psi_{r \alpha \sigma}^\dagger
\partial_x \psi_{r \alpha \sigma},
\label{eq:hkinetic}
\end{equation}
where the band shift and Zeeman term are left implicit. The operator
$\psi$ denotes the annihilation of a fermion and $\sigma=\pm$
correspond to electronic spin components. For large but realizable
tubes of $R = 3.39$ nm and experimentally accessible field strengths
$B=6.4$ T and $E=0.02$ V/nm, we find $v_R = 0.89 v_F$ and $v_L = 0.76
v_F$, yielding a pronounced asymmetry in right- and left-moving
velocities.

Transverse fields significantly alter not only the low-energy band
structure but also the nature of Coulomb interactions within the
tube.  The usual forward scattering contribution, $H_{C1} =
\frac{1}{2} \int dx  \  V \rho_{c+}^2(x)  $, remains and continues to
dominate for larger tubes ($n \gtrsim 10$)~\cite{kbf}. Here
$\rho_{c+}(x)= \rho_{R,c+} + \rho_{L,c+}$ with $\rho_{r,c+} =
\sum_{\alpha \sigma} \psi^\dagger_{r \alpha \sigma} \psi_{r \alpha
\sigma}$ is the net charge density and $V\approx \frac{2 e^2}{\kappa}
\ln \left( \frac{L}{2 \pi R} \right)$ is the effective
one-dimensional interaction~\cite{egger,us}, where $e$ is the charge
of an electron, $\kappa$ is the dielectric constant, and $L$ is the
length of the tube.  More interestingly, a second contribution
becomes manifest due to the uneven distribution of charge in the
circumferential direction induced by the presence of the applied
fields. A careful accounting of circumferential modes shows that this
additional effective Coulomb term along the tube has the form
\begin{equation}
H_{C2} = \frac{\pi \lambda}{4} \int dx \  \left( \rho_{R,c-} + \rho_{L,c-} \right) \left( \rho_{R,c+} - \rho_{L,c+} \right)
\label{eq:HC2}
\end{equation}
where $\lambda = 4 e^2 b u h /  \pi \kappa$ and $h$ is a numerical
factor which depends on the details of the geometry of the underlying
graphene lattice; for $n \gtrsim 20$ we have $ h \approx \left( 1.46
n - 4.60 \right)$~\cite{us}. Here, $\rho_{r,c-} = \sum_{\alpha
\sigma} \alpha \psi^\dagger_{r \alpha \sigma} \psi_{r \alpha \sigma}$
is the difference in charge density between the two bands.  The
presence of this interaction term has two origins: (i) the magnetic
field couples to the crystal momentum, and thus the Fermi points,
giving rise to the contribution involving the charge in the $c-$
sector and (ii) the electric field couples to the net charge
imbalance in bonding and anti-bonding states, or alternatively, to
the density difference in the right- and left-movers in the $c+$
sector. This interaction, which is key to the physics studied here,
goes a step beyond the usual spin-charge-separating Coulomb term by
directly coupling to the band degrees of freedom.

As with the field-free case, the asymmetric gapless modes and Coulomb
interactions can be studied via
bosonization~\cite{fernandez,naon,trushin,gil}, which renders the total
Hamiltonian for the tube to be quadratic. The one-dimensional fermionic
operators can be bosonized as
\begin{equation}
\psi_{r \alpha \sigma} = \frac{\eta_{r \alpha \sigma}}{\sqrt{2 \pi a_{c}}} \exp \left[ i \alpha k_F x + i \varphi_{r \alpha \sigma} \right],
\end{equation}
where the $\eta_{r \alpha \sigma}$'s are the so-called Klein factors
which enforce anticommutation relations between different
channels~\cite{egger}. The chiral bosonic fields satisfy the
commutation relations
$[\varphi_{r\alpha\sigma}(x),\varphi_{r'\alpha'\sigma'}(x')]=-i\pi
r\delta_{rr'}\delta_{\alpha \alpha'}\delta_{\sigma\sigma'}{\rm
sgn}(x-x')$. The density associated with each sector takes the form
$\rho_{r \alpha \sigma} = r
\partial_x \varphi_{r \alpha \sigma} / 2 \pi$. Given the asymmetric dispersion and current dependent interaction,
 we explicitly employ these chiral fields to ensure that this
algebra is preserved. It is convenient to introduce a spin and channel
decomposition for the chiral fields, $\varphi_{r, \alpha \sigma} = \frac{1}{2}
\left( \varphi_{r, c+} +  \alpha \varphi_{r,c-} + \sigma \varphi_{r,s+} +
\alpha \sigma \varphi_{r,s-} \right)$. The total kinetic energy takes the
bosonized form
\begin{equation}
H_0 = \frac{1}{4 \pi} \sum_{a = c/s\pm} \int dx \left[v_R \left( \partial_x \varphi_{R,a} \right)^2 + v_L \left( \partial_x \varphi_{L,a} \right)^2 \right],
\end{equation}
and the total interaction term, $H_{int} \equiv H_{C1} + H_{C2}$,
takes the form
\begin{eqnarray}
&& \!\!\!\!\!\!H_{int} = \frac{V}{2 \pi^2} \int dx  \left( \partial_x \varphi_{R, c+} - \partial_x \varphi_{L, c+} \right)^2 \\
          &&\  + \frac{\lambda}{4 \pi} \int dx  \left( \partial_x \varphi_{R,c+} + \partial_x \varphi_{L,c+} \right) \left( \partial_x \varphi_{R, c-} - \partial_x \varphi_{L,c-}
          \right).\nonumber
\label{eq:Hint}
\end{eqnarray}

As discussed in previous work on asymmetric
bosonization~\cite{fernandez,naon,gil,trushin}, in diagonalizing the
full Hamiltonian $H = H_0 + H_{int}$, care needs to be taken to
ensure that the diagonal basis  preserves the chiral commutation
rules respected by the original fields, $\varphi_{r\alpha\sigma}$.
Here, we circumvent unwieldy manipulations involving Bogoliubov
transformations by employing the trick of converting left-handed
fields to right-handed fields via the transformation $\varphi_{L a}
\mapsto i \varphi_{L a}$, which then allows for standard rotations.
The resultant plasmon modes for the coupled charge sectors $c\pm$
move with four different velocities
\begin{eqnarray}
\!\!\!\!\!&&v_{R/L,1/2}   =  \pm \epsilon + v_{1/2}, \label{eq:vBE} \\
\!\!\!\!\!&&v_{1/2}   =  \frac{v}{\sqrt{2} K_{c+}} \Bigg[ 1\! +\! K_{c+}^2 \pm
\sqrt{ \left(1\!-\!K_{c+}^2 \right)^2 +  \left( \frac{2K_{c+}\lambda}{ \hbar
v} \right)^2} \Bigg]^{\frac{1}{2}},\nonumber
\end{eqnarray}
where  $v = (v_R + v_L)/2$ is the average velocity of the
non-interacting fermions, $\epsilon = (v_R - v_L)/2$ the asymmetry,
and $K_{c+} \equiv 1/\sqrt{1+ 4 V/\pi \hbar v}$ is the standard
Luttinger parameter. The spin sectors $s\pm$ each retain their band
structure velocities $v_{R/L}$.

The existence of the two different plasmon velocities $v_{1/2}$
demonstrates that transverse fields can access the band sector, which
is normally resilient to any bulk coupling. Thus far, momentum
resolved bulk tunneling experiments on quantum wires~\cite{Yacoby05}
have revealed two distinct plasmon dispersions corresponding to spin
and charge modes; analogous experiments on nanotubes in transverse
fields ought to unveil the three pairs of velocities predicted above.
For example, for a tube $R = 3.39$ nm and experimentally accessible
field strengths $B=6.4$ T and $E=0.02$ V/nm, the plasmon velocities
are given by $v_{R,1} = 4.07 v_F$, $v_{L,1} = 3.94 v_F$, $v_{R,2} =
0.29 v_F$, and $v_{L,2} = 0.16 v_F$ for the charge sectors and $v_R =
0.89 v_F$ and $v_L = 0.76 v_F$ for the spin sectors. The observation
of three distinct modes would reflect complete splintering of the
nanotube electron into its spin, charge and band sectors.

The effects of transverse fields on the low-energy nanotube physics,
particularly in the band degree of freedom, are prominently
manifested in physical observables. The tunneling density of states,
an ubiquitous, experimentally accessible quantity, retains its the
power-law form $\chi(E) \sim E^{\beta}$ characteristic of Luttinger
liquids but exhibits a modified exponent. Using standard
procedures~\cite{giamarchi} that now account for field-dependent
effects, we find that the tunneling exponent, to lowest order in
$\lambda$, is given by
\begin{eqnarray}
\beta_{r\alpha} = \frac{1}{8} \left( \frac{1}{K_{c+}} + K_{c+} -2 \right) -
\frac{\lambda}{4 \hbar v} r \alpha \left(\frac{1-K_{c+}}{1+K_{c+}} \right).
\label{eq:dos}
\end{eqnarray}
The first part of the exponent, also present in the field-free case,
reflects the suppression in tunneling due to interactions (where
$K_{c+}$ is now tunable). The second part, which depends on the
tunable coupling $\lambda$ of $H_{C2}$, further suppresses or
enhances tunneling depending on the sign of $r \alpha$. As an
estimate, for a $R = 2.4$ nm tube in a $5.9$ T B-field and $0.023$
V/nm E-field with a field-free value of $K_{c+} = 0.22$, we have
distinctly different exponents $\beta_{R +} = \beta_{L - } = 0.32$
and $\beta_{R - } = \beta_{L +} = 0.39$. The form of
Eq.~(\ref{eq:dos}) reflects band (valley) selection; for example, a
right-moving electron would preferentially tunnel into the $\alpha =
+$ Fermi point for the field configuration shown in
Fig.~\ref{fig:setup}

A physical consequence of transverse fields yielding such a
band-dependent exponents would be the presence of two different
power-law contributions to the non-Ohmic conductance of the
nanotube~\cite{yao}. On a related note, the presence of an impurity
would distinguish these exponents; conductance properties would be
sensitive to whether or not an electron impinging on the impurity
switched band index. We note here that the related charge current,
obtained from the continuity equation, has an atypical form, $J_{c+}
= v_R \rho_{R,c+} - v_L \rho_{L,c+} + \lambda \left( \rho_{R,c-} +
\rho_{L,c-} \right)$. Band-dependent effects similar to those in the
density-of-states exponent would be manifest in any susceptibilities
involving the sublattice degree of freedom. Tendencies towards
different orderings, such as charge or spin density waves, which are
governed by associated susceptibilities, would in turn reflect band
dependence. For instance, charge density waves have contributions
from operators
$\hat{O}_{CDW\pi}^{\pm}\sim\sum_{r\alpha\sigma}r\psi^\dagger_{r\alpha\sigma}\psi_{-r\pm\alpha\sigma}$,
reflecting like ($+$) and staggered ($-$) band correlations at A and
B sublattice sites. While these operators both come on an equal
footing in the field-free case, we find that fields render
$\hat{O}_{CDW\pi}^-$ more relevant in the renormalization group
sense; fields can thus discriminate between two types of
band-dependent ordering at the sublattice level.

In the quantum dot limit achieved by high resistance contacts or
sufficiently low temperatures, we find that transverse fields enable
controlled tuning of Coulomb blockade physics and nanotube quantum
states. Here, as in previous treatments~\cite{kbf,fabrizio,eggert},
we describe the dot as a finite-sized version of the net nanotube
Hamiltonian and focus on the topological sectors as relevant for
standard quantum dot experiments involving adiabatic tuning. In the
presence of a gate voltage $V_G$, the resultant Hamiltonian
characterized by quantum numbers $N_a$ for each topological sector is
given by
\begin{eqnarray}
H_T = \sum_{a=c/s\pm} \frac{\epsilon_a}{8}  N_a^2 &-& \mu N_{c+} - \frac{\pi\lambda}{4L} \left( \frac{v_R - v_L}{v_R + v_L} \right) N_{c+} N_{c-} \nonumber\\
       &+& \frac{1}{2} \Delta_{b}N_{c-} - \Delta_Z N_{s+},
\label{eq:Ham_qtmdot}
\end{eqnarray}
where $\epsilon_a = \epsilon_0 + 4 E_a$, $\epsilon_0=\frac{2\pi \hbar
v_Rv_L}{L(v_R+v_L)}$ and $E_a$ is the interaction strength in a given
mode. Here, $E_{c+} = V/L$, $E_{c-} = - \pi \lambda^2/ 4 \hbar v L$
and $E_{s\pm}=0$. The term $\mu$ is proportional to $e V_G$, and
$\Delta_Z = \mu_B B$ accounts for the Zeeman splitting.  The coupling
between the $c+$ and $c-$ sectors arises due to the the interaction
$H_{C2}$ (Eq.~(\ref{eq:HC2})) and the no-current boundary conditions
at the tube ends. The band splitting $\Delta_b$ also depends on
boundary conditions; in the absence of Fermi point mixing at the tube
ends, it reduces to the band structure Fermi point mismatch
$\Delta_F$~\cite{us}. As typical parameter values, for a 1
$\mu$m-tube of radius 3.39 nm, and magnetic and electric field
strengths of 4.51 T and 0.026 V/nm, we find $\epsilon_0 =1.32 $ meV,
$E_{c+}/\epsilon_0 = 6.14$, $E_{c-}/\epsilon_0 = -0.23$,
$\Delta_Z/\epsilon_0 = 0.20$, and $\lambda / L \epsilon_0 = 0.24$.

\begin{figure}
    \includegraphics[bb=165 25 525 525,height=57mm]{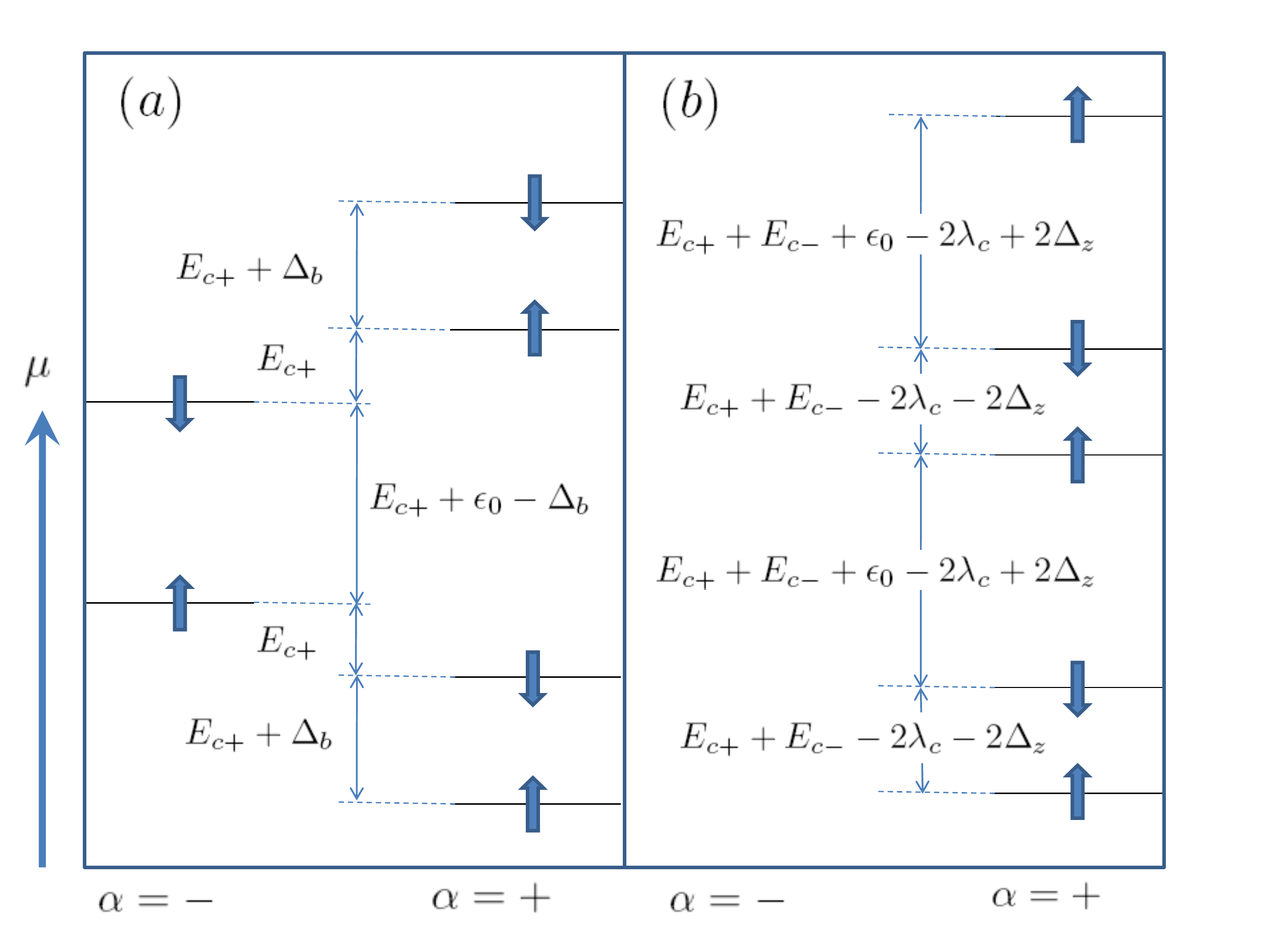}
     \caption{(Color online) Examples of shell-filling. Figure indicates the order, $\alpha$-point, and spin of tunneled electron with increasing chemical
     potential $\mu$ (not drawn to scale). {\bf (a)} Filling order for a tube in the absence
     of fields and with $\Delta_b\neq 0$. {\bf (b)} Shell-filling for $N_{c+}=N_{c-}$ exhibits 2-fold periodicity and complete polarization into band $\alpha = +$.
     This condition is approximately met for a 1 $\mu$m-tube of radius 3.39 nm in the presence of magnetic and electric fields 4.51
T and 0.0255 V/nm, respectively (see text for specific values of the Coulomb blockade parameters for this case and other details).
     \label{fig:CB}}
\end{figure}

By virtue of their field dependence, several parameters in $H_T$ can
be varied to access a wide variety of shell filling configurations of
the nanotube quantum dot in Coulomb blockade experiments. The
configurations correspond to sets of electron occupation numbers
which minimize the energy associated with
Eq.~(\ref{eq:Ham_qtmdot})~\cite{kbf,us}. While the associated
parameter space is too extensive for an exhaustive study, a few
salient characteristics of shell-filling are as follows. (i) In
actual experiments~\cite{persistent,bockrath}, the band degeneracy is
often naturally broken due to physical attributes such as the
confining potential created by the leads, yielding patterns such as
in Fig.~\ref{fig:CB}a. Here, the field dependence of $\Delta_{b}$
enables controlled tuning of band degeneracy breaking as well as
probing the extent of Fermi point mixing at the tube
ends~\cite{mccann,us}. (ii) A four-fold periodicity has been observed
in some experiments reflecting the band and spin degeneracy of the
tube~\cite{bockrath}. (iii) As a direct demonstration of band tuning,
the parameter values quoted above yield the two-fold periodic
shell-filling pattern in Fig.~\ref{fig:CB}b. In particular, these
values are chosen such that Eq.~(\ref{eq:Ham_qtmdot}) is minimized by
the condition $N_{c+} = N_{c-}$, entirely restricting the tunneling
into a given $\alpha$-point. We remark that this condition requires a
fine-tuning of fields, and also that other field values can even
render $\epsilon_{c-}$ negative, resulting in an instability towards
complete polarization into one band. (iv) Periodicity can also be
entirely obliterated by choosing an irrational ratio between two of
the relevant shell-filling parameters. As demonstrated above,
transverse fields provide a precise means of preparing and
manipulating the electronic spin and band quantum numbers of the
nanotube quantum dot.

In conclusion, we have presented transverse fields as powerful probes
to access and explore a rich range of physics in armchair SWNTs that
directly addresses the band degree of freedom. We have found that
these fields induce an unusual interaction that couples the charge
and band sectors. We have predicted that signatures of these field
effects will be apparent in a variety of measurements including those
probing Luttinger liquid parameters and plasmon structure, tunneling
density of states and impurity scattering.  We have shown that
transverse fields can be used for controlled manipulation of nanotube
quantum dot states, making this proposed setup a potential building
block for nanoscale quantum devices.

We would like to thank N. Mason and M. Stone for their perceptive
comments. This work is supported by NSF, Grant No. DMR09-06521 (W. D.
and S. V.), NSERC and MITACS (T.-C. W.), and UNLP and CONICET (V. F.)


\begin{thebibliography}{99}
\bibitem{gap} For non-perpendicular fields oriented at an angle
    $\chi$, a relevant coupling between right- and left-moving
    electrons opens a bare gap of order $t b u (2 \sqrt{3}\cos
    \frac{\pi}{3n}) \left| \cos \chi \right|/(1 + 2
    \cos{\frac{\pi}{3n}})$. This gap however is rendered irrelevant
    by interaction effects.
\end{thebibliography}
\end{document}